# Low Cost Thermopile Detectors for THz Imaging and Sensing

F. Voltolina[a], A. Tredicucci[b] and P. Haring Bolivar[a]
[a] Institute HQE, University of Siegen, Hölderlinstr. 3, 57068 Siegen, Germany
[b] NEST-INFM and Scuola Normale Superiore, Piazza dei Cavalieri 7, 56126 Pisa, Italy

*Abstract*—We evaluate 2 thermopile sensors as convenient detector options for a modern integrated THz imaging system. Both a commercial single pixel device and a 32x32 pixel FPA are utilized in a simple setup with a single optical element to achieve an outstanding resolution and SNR performance. Either a cryogenically cooled quantum cascade laser (QCL) or a room temperature electronic THz multiplier chain is used as source.

## I. Introduction and Background

THE thermopile is the first practical infrared detector providing an electrical output signal: its invention and initial use dates back to 1820 with the physicists Oersted, Fourier, Nobili and Melloni. Despite its current availability as a mass produced device, often embedded in low cost consumer applications, to the best of our knowledge it received just a marginal attention in the THz community.

While in the medium infrared its application ranges from remote sensing thermometers to satellite attitude sensors, it is typically limited to the role of wide area power meter detector in the THz range. We investigated the single pixel TPS334 from Perkin Elmer[1] and the AXT100 camera from Ann Arbor Sensor Systems LLC[2], both designed for the mid infrared, to determine their suitability for a general THz imaging system.

## II. TPS 334 Single Pixel Detector: Results

The TPS334 single pixel detector is characterized by a sensitive area of 0.7x0.7 mm$^2$ and a responsivity of 55 V/W with a NEP of 0.64 nW/√Hz, without window.

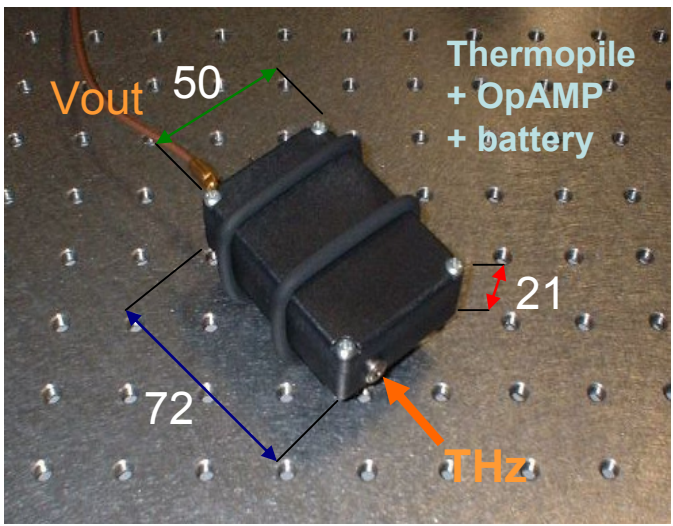

Fig. 1: Picture of our compact integrated detector.

The -3dB point in the frequency response is reached at 10 Hz, making its use attractive in a THz imaging system that has to scan a limited amount of sites per second (typically less than 100 sites in total, for the intended embodiment).

Our core application is a DNA biochip reader system[3] that inspired the development of a compact integrated detector realized around the TPS334 thermopile chip and a battery operated instrumentation amplifier chip (see Fig. 1).

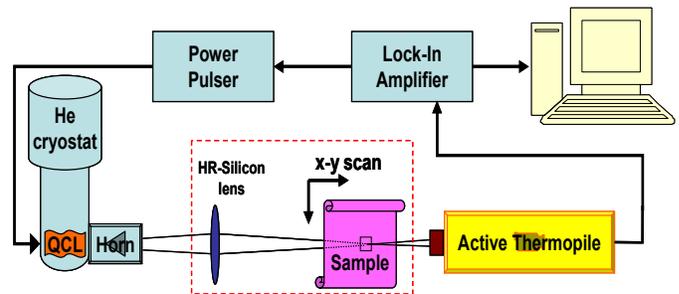

Fig. 2: Schematic drawing of the DNA biochip reader system.

The complete imaging system (see Fig. 2) uses a 2.8 THz QCL laser[4] cooled by a closed cycle Helium cryostat (model CCS-100/204 from Janis Research), a single high resistivity silicon lens (Focal length = 25 mm.) and a Newport 2-axes high resolution motor stage.

A simple test phantom was prepared, using a mail envelope filled with different materials known to cover a broad absorption range at THz frequencies, and imaged (see Fig. 3).

Images taken at 100μm pixel step size drastically outperformed that acquired with a Golay detector and even 250μm features are well resolved according to transmission imaging of a standard USAF resolution target (chrome on fused silica).

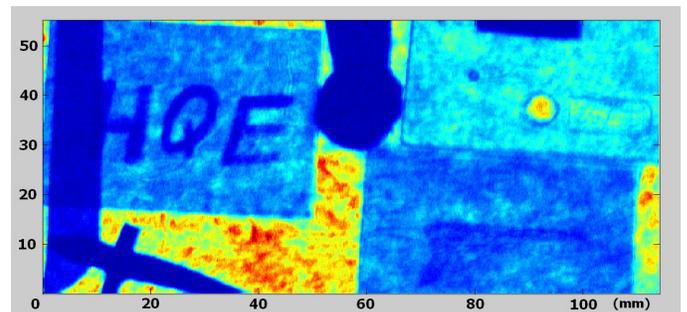

Fig. 3: Transmission image of an envelope at 2.8 THz.

Real dimensions of this target are 15x14 mm and the magnified part shows the good quality in rendering both regions with 2 and 2.24 line pair/mm (see Fig. 4). The advantages of the Golay cell in both sensitivity and NEP values vanishes in practical use due to the enormous increment in dynamic range provided by the linear operation of the thermopile up to 100 mW of incident power.

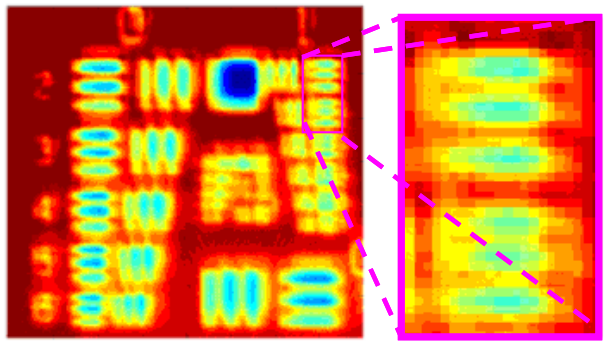

Fig. 4: Standard USAF resolution target: 2 and 2.24 line pair/mm.

At the power level available from the modern QCL sources, the system SNR is >50dB without the use of a lock-in amplifier. Higher SNR can be achieved with a lock-in amplifier, which can be additionally tuned to the cryostat vibrations in order to eliminate artifacts due to standing waves effects. This enhanced setup has immediately shown drastically minimized vibration-related issues and made the quest to achieve almost diffraction-limited performance as simple as to find the correct positioning in the setup of a single lens: an HR silicon element of 25mm diameter and focal length.

III. AXT 100 CAMERA: RESULTS

The AXT100 camera with custom window and modified electronic is used in an alternative imaging system. We performed real time beam profiling and imaging using either QCL or an Amplified Multiplied Source at 600 GHz from RPG Radiometer Physics (see Fig. 5). The detector array consists of 32x32 pixels that have an active area size of 300x300 μm and a responsivity of 90 V/W with a NEP of 0.49 nW/√Hz, without window.

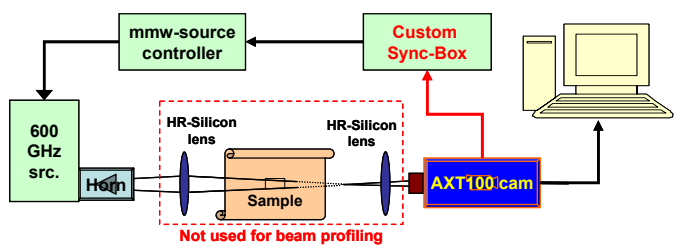

Fig. 5: Schematic drawing of the biochip parallel readout system.

Due to the constant response of the thermopile absorbers from mid infrared up to a wavelength of at least 100μm, there is a limited increase of optical NEP when operating the camera far away from the array design frequency.

Hence, this camera provides a valid alternative to more expensive FPA microbolometer cameras[5] (SCC500, BAE Systems) when operating with long wavelength THz radiation, between 600 GHz and 3 THz, and modest power levels. This is because a MIR imager incur in a drastic decrease in response when used in the FIR region, due to inefficient absorbers and sub-wavelength pixel size, while a device like the AXT 100 maintains its rated responsivity below the 1 THz mark by design. In this case the main limiting factor become the window protecting the sensor array with its negative influence in terms of reflection and transmission losses.

In the particular case of the AXT 100 a basic silicon low resistivity window can be swapped for a high resistivity one or more drastically for a high density polyethylene sheet.

Still, the high level of MIR background radiation is typically overwhelming the FIR we want to image, due to the blackbody radiation of any object at room temperature happening to be in the field of view.

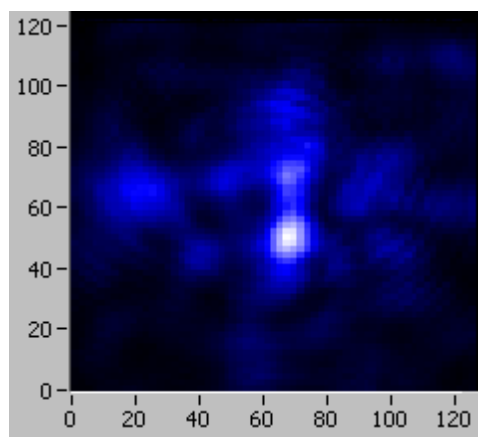

Fig. 6: Spatially upsampled beam profile of a 3.3 THz QCL.

A definitive modification to enable practical application of the AXT 100 as a THz imager is the invention of a custom sync-box, connected to the camera via an optical link, to enable a synchronous 'blinking' of the THz source with the frame rate of the camera. At that point it is easy to apply a differential scheme to subtract the background and obtain a nice THz image. Early test, without optics, allowed real time (15 fps) beam profiling of 2.8 and 3.3 THz QCLs (see Fig. 6).

Real time imaging samples and additional application examples and characterizations of both imaging systems will be shown.

REFERENCES

[1] Perkin Elmer, "TPS334 thermopile sensor product specification"
[2] Ann Arbor Sensor Systems LLC, "AXT100 Thermal Imaging Camera"
[3] C. Debus, F. Voltolina and P. Haring Bolivar, "Towards cost-efficient THz biochip technologies" in *Antennas and Propagation International Symposium, 2007 IEEE*, June 2007, pp. 3380-3383.
[4] R. Köhler, A. Tredicucci, F. Beltram, H. E. Beere, E. H. Linfield, A. G. Davies, D. A. Ritchie, R. C. Iotti, and F. Rossi, "Terahertz semiconductor-.heterostructure laser" Nature 417, 156–159, 2002.
[5] Alan W. M. Lee, Qi Qin, Sushil Kumar, Benjamin S. Williams, Qing Hu, John L. Reno, "Real-time terahertz imaging over a standoff distance (> 25 meters)" APL, October 2006, Vol 89, Issue 14.